\documentclass[pra,reprint,floatfix,longbibliography,superscriptaddress]{revtex4-1} 
\usepackage{amsmath}  
\usepackage{amsfonts} 
\usepackage{graphicx} 
\usepackage{bm}
\usepackage[usenames,dvipsnames]{color}
\usepackage{verbatim}
\usepackage[T1]{fontenc}
\usepackage{MnSymbol}
\usepackage{wasysym}
\usepackage{soul}
\usepackage{anyfontsize}
\usepackage{subfig} 
\usepackage{ upgreek }

\def\be{\begin{eqnarray}}
\def\ee{\end{eqnarray}}

\def\He3{$^3$He}

\def\He3{$^3$He}

\RequirePackage{lineno}

\begin{document} 

\setpagewiselinenumbers

\modulolinenumbers[5]

\title{Continuous Comagnetometry using Transversely Polarized Xe Isotopes}
\def\wisc{Department of Physics, University of Wisconsin-Madison, Madison, Wisconsin 53706, USA}
\author{D. Thrasher}
\affiliation{\wisc}
\author{ S. Sorensen}
\affiliation{\wisc}
\author{ J. Weber}
\affiliation{Natural Sciences Department, Parkland College, Champaign, IL 61821, USA}
\author{M. Bulatowicz}
\affiliation{\wisc}
\author{ A. Korver}
\affiliation{\wisc}
\author{ M. Larsen}
\affiliation{Northrop Grumman, Advanced Position, Navigation and Timing Systems, Woodland Hills, CA 91367, USA}
\author{ T. G. Walker}
\affiliation{\wisc}
\email{tgwalker@wisc.edu}

\date{\today}

\begin{abstract}{
We demonstrate a transversely polarized spin-exchange pumped noble gas comagnetometer which suppresses systematic errors from longitudinal polarization. Rb atoms as well as $^{131}$Xe and $^{129}$Xe nuclei are simultaneously polarized perpendicular to a pulsed bias field. Both Xe isotopes' nuclear magnetic resonance conditions are simultaneously satisfied by frequency modulation of the pulse repetition rate. The Rb atoms detect the Xe precession. We highlight the importance of magnetometer phase shifts when performing comagnetometry. For detection of non-magnetic spin-dependent interactions the sensing bandwidth is 1 Hz, the white-noise level is 7 $\mu$Hz /$\sqrt{\text{Hz}}$, and the bias instability is $\approx1$ $\mu$Hz. 
}\end{abstract}

\maketitle

Spin-exchange (SE) pumped comagnetometers~\cite{Limes2018,Walker2016} utilize co-located ensembles of spin-polarized alkali-metal atoms and noble gas nuclei~\cite{Walker1997} to suppress magnetic field noise. Such devices have been used to place upper bounds on spin-mass couplings~\cite{Bulatowicz2013,Lee2018}, Lorentz violations~\cite{Allmendinger2014,Romalis2014,Smiciklas2011,Brown2010,Glenday2008}, and atomic electric dipole moments~\cite{Rosenberry2001,Allmendinger2019,Sachdeva2019}, and for the measurement of inertial rotation~\cite{Walker2016,Kornack2005,Jiang2018,Karwacki1980}. The fundamental uncertainty of a SE pumped comagnetometer's measure of inertial rotation scales favorably with sensor size compared to alternative technologies~\cite{Donley2010}.

Longitudinal SE fields are important sources of systematic uncertainty in devices which utilize the embedded alkali-metal atoms for high SNR detection. Consider an ensemble of two noble gas species ($a$ and $b$) which are spin-exchange optically pumped (SEOP) in a common magnetic field $B_{z}$ and are each subject to some spin-dependent interaction $X$. The Larmor resonance frequency of isotope $a$ can be written as~\cite{Walker2016,Limes2019,Terrano2019,Petrov2019}
\begin{equation}
\Omega^{a} = \gamma^{a}( B_{z}+b_S^a S_z+ b_b^a K_z^b) + X_z^{a},
\end{equation}
where $\gamma$ is the gyromagnetic ratio, $S$ and $K$ are the respective alkali-metal and noble gas polarizations, $z$ subscripts refer to the longitudinal components (\textit{i.e.}, parallel to the bias field direction), and $b^i_j$ is the SE coefficient characterizing the influence of $j$'s polarization on $i$. With knowledge of $\rho = \gamma^a$/$\gamma^b$, simultaneous measurement of $\Omega^a$ and $\Omega^b$ allows $B_z$ to be suppressed while sensitivity to $X_z^a$ and $X_z^b$ is retained~\cite{Chupp1988}.

\begin{figure}
\includegraphics[width = 8.6cm]{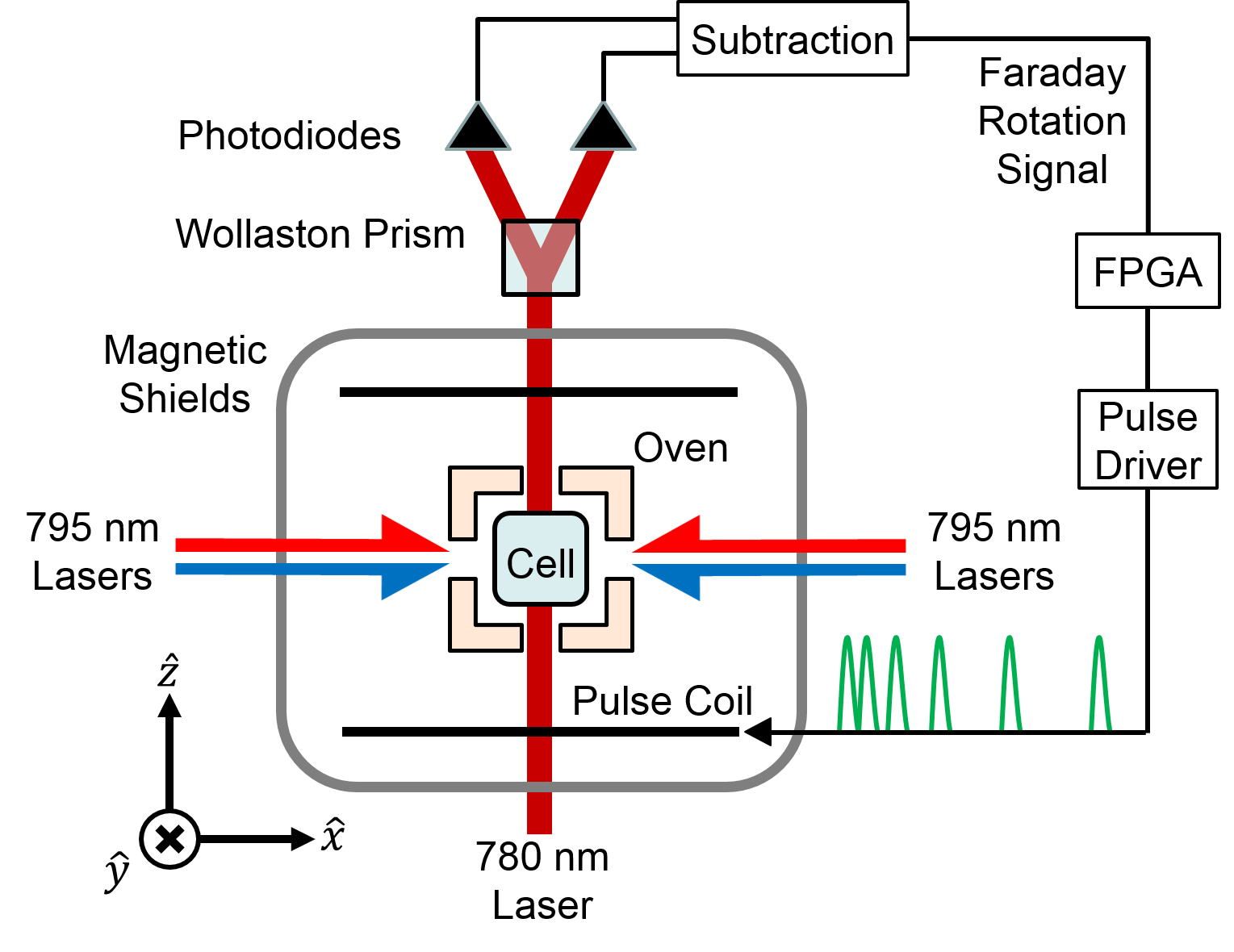}
\caption{Schematic of apparatus. Field shim coils are not shown. The green trace depicts the frequency modulated bias field pulses. }\label{schem}
\end{figure}

In this paper we demonstrate a SE pumped $^{131}$Xe-$^{129}$Xe comagnetometer which suppresses time-averaged $S_z$ and $K_z$ such that
\be\label{master}
\xi \equiv  {\rho \Omega^b - \Omega^a\over 1+\rho}  \approx {\rho X_z^b - X_z^a\over 1+\rho},
\ee
where the superscripts $a$ and $b$ refer to $^{129}$Xe and $^{131}$Xe, respectively, and $\rho = 3.373417(38)$~\cite{Makulski2015}. The comagnetometer relies on a dual-species version of synchronous SEOP~\cite{Korver2015}, wherein noble gas nuclei are continuously polarized transverse to a frequency modulated pulsed bias field. We demonstrate that the correlation between the frequency of precession of $^{131}$Xe and $^{129}$Xe is sufficient to resolve 7 $\mu$Hz /$\sqrt{\text{Hz}}$ of white frequency noise with a low frequency field noise suppression of $>10^3$. We also demonstrate the influence of alkali-metal magnetometer phase shifts on the field noise suppression of the comagnetometer. 

A schematic of the experimental setup is shown in Fig.~\ref{schem}. $^{85}$Rb atoms are optically pumped along $\hat{x}$, transverse to a pulsed bias field oriented along $\hat{z}$. The noble gas nuclei are polarized via SE collisions with the Rb atoms. The bias field is applied as a series of low duty cycle pulses (depicted in green), where each pulse produces a $2 \pi$ Larmor rotation of the Rb spins. Due to the $\sim10^3$-fold smaller magnetic moments of the Xe isotopes, they experience $\approx 2\pi$/$10^3$ radians of precession per pulse, approximating a continuous bias field of $B_{p}(t) =\omega_{p}(t) / \gamma^S$, where $\omega_{p}(t)$ is the repetition rate of the $2 \pi$ pulses and $\hbar \gamma^S=2\mu_B/(2I+1)$, where $I=5/2$ is the $^{85}$Rb nuclear spin and $\mu_B$ is the Bohr magneton. The magnetic resonance of each noble gas species is simultaneously excited by modulating $\omega_p$ at linear combinations of  $\Omega^a$ and $\Omega^b$.

The experimental apparatus is similar to Ref.~\cite{Korver2015}. Updates include an 8 mm cubic Pyrex cell filled with 40 Torr enriched Xe and 50 Torr N$_2$ with a hydride coating~\cite{Kwon1981}, and a custom-made pulse driver designed so that the voltage required to produce 2$\pi$ pulses is largely independent of pulse repetition rate.

The embedded Rb magnetometer, which is effectively at zero-field due to the low duty cycle nature of the bias field pulses~\cite{Korver2013}, continuously detects the noble gas SE field $b_K^S K_y$. The Faraday rotation of a linearly polarized probe laser propagating parallel to the bias field serves as a monitor of $S_z\propto K_y$. 
\begin{figure}
\includegraphics[width = 8.6cm]{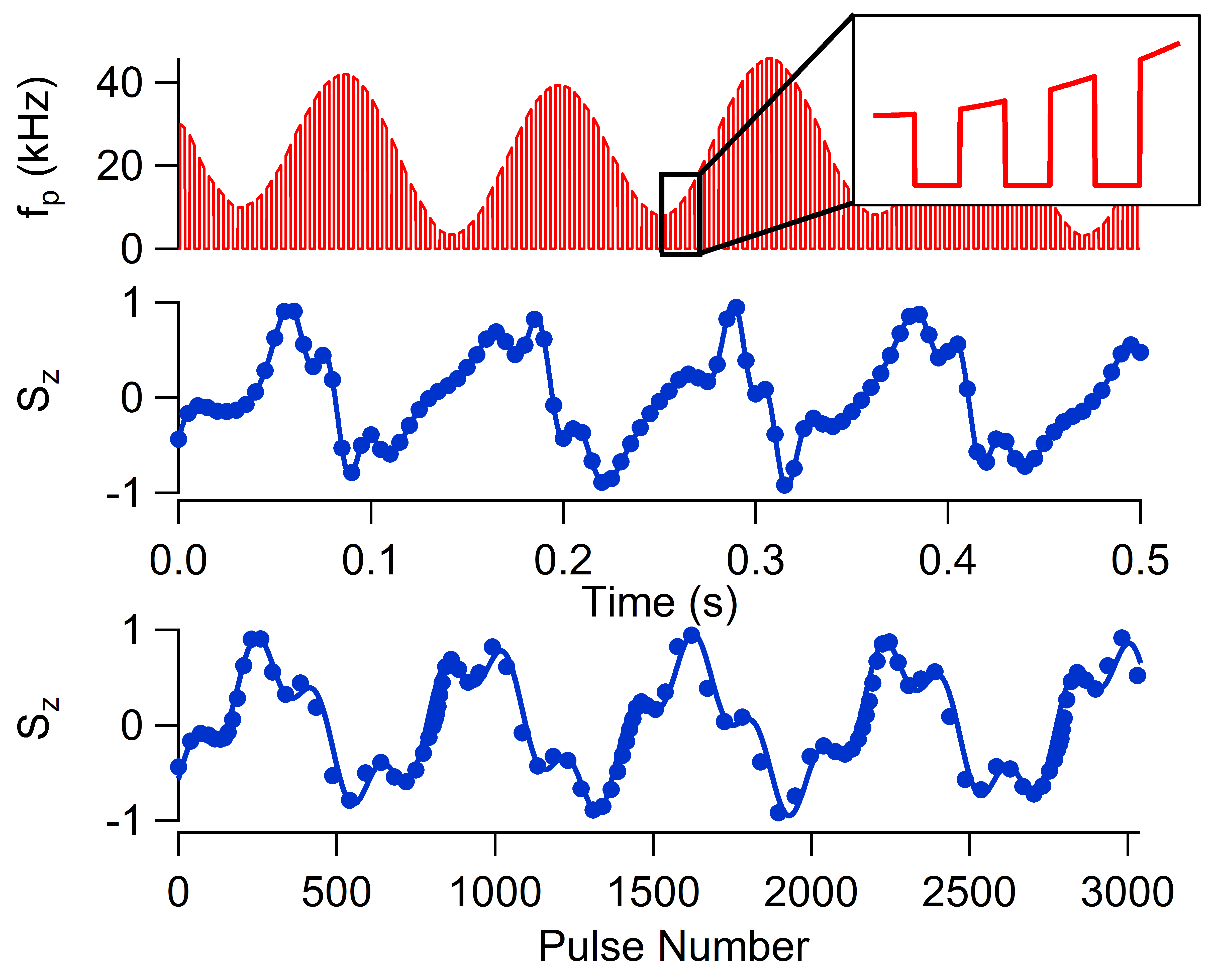}
\caption{Time dependence of bias pulse repetition rate $f_p$ (top) and normalized $S_z$ (middle) for $\Delta^a = \Delta^b \approx 0$. Corresponding pulse number dependence of normalized $S_z$ (bottom). Filled circles are measured data. Lines are theory fits to the data. Inset depicts greater time detail of the bias pulse gating.}\label{sigs}
\end{figure}
The phase of a rotating transverse field as measured by the Rb depends on the stray field $B_{z0}$. We write the transverse spin of the Rb valence electron as $S_+ = S_{\perp} e^{i\epsilon_z}$, where $\epsilon_z = \tan^{-1}({ S_y \over S_x}) \equiv \tan^{-1}({ B_{z0} \over B_w})$ is the magnetometer phase shift due to $B_{z0}$ and $B_w$ is the magnetic width of the magnetometer. The $S_z$ signal is digitally demodulated, and the pulse modulation frequencies are adjusted to maintain resonance for both Xe species. The Xe resonance frequencies determined by the feedback are recorded and used to compute $\xi$. We will show that including $\epsilon_z$'s contribution to $\xi$ is crucial for optimizing the comagnetometer's suppression of low frequency field noise. 

The optically pumped Rb atoms produce a SE field $b_S^K S_x$ experienced by the Xe which induces broadening of the NMR and the production of $K_z$. We take care to null transverse fields experienced by the Xe~\cite{Korver2015}.  Under these conditions, the transverse components $K_+=K_x+iK_y$ of the nuclear spin polarization for each noble gas species obey
\be\label{BE1}
{dK_+\over dt}=-(\mp i\Omega+\Gamma_2)K_+ +\Gamma_{S}^K S_+,
\ee
where $\Omega=\gamma B_z+ X_z$ is the Larmor resonance frequency, arising from both magnetic field $B_z$ and spin-dependent phenomena $X_z$.  The SE rate constant is $\Gamma_{S}^K$, and the transverse relaxation rate is $\Gamma_2$. The sign in front of $\Omega$ encodes the direction of precession (top is $^{129}$Xe, bottom is $^{131}$Xe). We assume that $^{131}$Xe frequency shifts due to quadrupole interactions are independent of $B_z$ and are included in $X_z^b$.

Multiple noble gas species can be simultaneously excited via SE collisions with the Rb by modulating either $S_+$~\cite{Korver2015} or $B_z$ in Eq.~\ref{BE1}. Here we choose to modulate $B_z$ by modulating $\omega_{p}$. The Rb magnetometer is far less sensitive to changes in $\omega_{p}$ than to changes in DC fields.

The magnetic field component along $\hat{z}$ is $B_z=B_{z0}+B_p(t)$, which consists of the stray field $B_{z0}$ and the field from the $2\pi$ pulses $B_p(t) = B_{p0}+B_m(t)$, where $B_{p0}$ is the DC component and $B_m$ is the AC component of the pulsed field. Since each pulse produces $2\pi$ precession of the Rb atoms, the Rb atoms primarily experience $B_{z0}$, while the Xe nuclei experience both $B_{z0}$ and $B_p$. In practice, we find that the magnetometer gain varies up to a factor of two as the pulse repetition rate is modulated (the magnetometer would not function if $B_{z0}$ were modulated instead). We avoid such gain modulation by gating the 2$\pi$ pulses and only recording the Faraday signal when the pulses are off and the magnetometer gain has stabilized. The $2\pi$ pulse repetition rate (depicted in Fig.~\ref{sigs}) is modulated as
\be \label{myeq}
\omega_p(t) = \omega_{p0}g(t)(1+ b_1 \cos(\omega_1t)+b_2 \cos(\omega_2 t)),
\ee
where $g(t) = (\rm{sign}(\cos(\omega_3 t))+1)$ is the time dependence of the gating, $\omega_1 = \omega_d^{b}$ and $\omega_2 = \omega_d^{a} - 3\omega_d^{b}$ determine the Xe drive frequencies, and $b_1$ and $b_2$ set the depth of modulation.


We use a phasor representation $K_+ = K_{\perp} e^{ \pm i\phi}$ for the nuclear spin polarization. In the experiment we measure the difference $\delta = \phi - \alpha$ between the instantaneous Xe phase $\phi$ and a reference phase $\alpha = \int(\omega_d + \gamma B_m) dt$ which is the phase the Xe would have if the only fields present were the pulsing fields and if $\Delta \equiv\omega_d - \Omega_0=0$ with $\Omega_0=\gamma(B_{z0}+B_{p0})+X_z$. To first order in $\delta$ and $\epsilon_z$, the imaginary part of Eq.~\ref{BE1} is
\be\label{6b}
{d\delta\over dt} =-\Delta-\Gamma_2(\delta \mp \epsilon_z),
\ee
and the time average of the real part is $K_{\perp}= \Gamma_S^K S_{\perp} C$/$\Gamma_2$, where $C$ is the time average of $\cos(\alpha)$. Note that the precession direction makes the sign in front of $\epsilon_z$ isotope dependent.  Performing a Fourier transform gives
\be\label{openloop}
\tilde{\delta} = -{\tilde{\Delta} \mp \Gamma_2 \tilde{\epsilon}_z \over i\omega +\Gamma_2}.
\ee

We detect the z-component of the Rb spin,
\begin{eqnarray}
S_z = {1 \over B_w} (b_{a}^S \mathbf{K^a} \times \mathbf{S}+b_{b}^S \mathbf{K^b} \times \mathbf{S})_z \nonumber
\\
= A^{a}_{\perp} \sin(\alpha^a+\delta^{a} - \epsilon_z)+A^{b}_{\perp} \sin(\alpha^b+\delta^{b} + \epsilon_z).
\end{eqnarray}
This assumes that $\mathbf{K}$ precesses slowly enough that $S_z$ adiabatically responds. It also assumes negligible back polarization from the Xe to the Rb~\cite{Bhaskar80b,Limes2018}. We see that the detected phase of $S_z$ depends on the Xe phases as well as the magnetometer phase.

Figure~\ref{sigs} shows $S_z$ for $\Delta^a = \Delta^b \approx 0$ as a function of both time and pulse number. Although the signal exhibits a complicated time dependence, its dependence on pulse number (which is equivalent to its dependence on $\alpha$ if $\Delta =0$) is a simple superposition of sine waves. 

We extracted the precession phase of each isotope as measured by the Rb by demodulating $S_z$ with $\cos(\alpha)$ of each isotope. Since $S_z$ was sampled at $\omega_3$, we applied an anti-aliasing filter before approximating the time average by a moving average over $N$ data points, where $N$ was chosen to remove the high frequency residuals of the demodulation.
\begin{figure}
\includegraphics[width = 8.6cm]{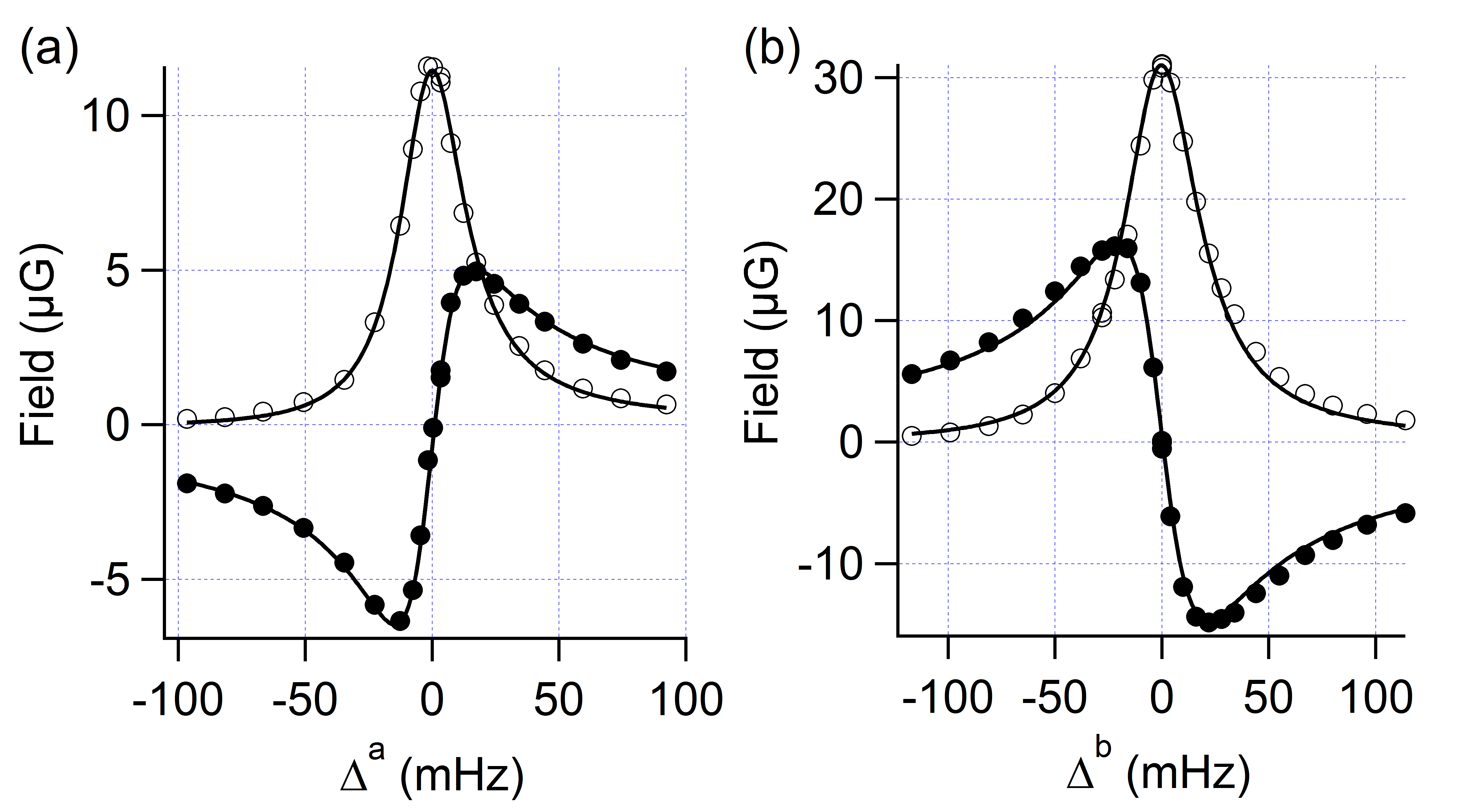}
\caption{Xe NMR Lineshapes. (a) $b_a^S K_{\perp}^a$ vs $\Delta^a$. (b) $b_b^S K_{\perp}^b$ vs $\Delta^b$.  Filled circles are $K_x$ and open circles are $K_y$. Lines are Lorentzian fits to the data.}\label{nmr}
\end{figure}
Figure~\ref{nmr} shows the demodulated $S_z$ for each isotope as $\omega_d$ is scanned, giving a familiar NMR resonance shape. The data were acquired with one isotope driven on resonance while the other isotope's detuning was varied. For the field modulation, we set $\omega_3 = 2\pi\times200$ Hz, $b_1=0.73$ and $b_2=0.15$, and $\omega_{p0}\approx 2\pi\times13.2$ kHz, resulting in average precession frequencies of $\approx 33.3$ Hz and $\approx 9.9$ Hz for $^{129}$Xe and $^{131}$Xe, respectively. The moving average filter was $N$=162. With these settings, the amplitude of $B_m$ was about $10B_w$ and the detection bandwidth was 1 Hz.

For $^{131}$Xe we realize $b_b^S K^b_{\perp}\approx$ 30 $\mu$G (corresponding to 0.1\% polarization) and a linewidth of $21.2(3)$ mHz. For $^{129}$Xe we realize $b_a^S K^a_{\perp}\approx$ 10 $\mu$G (0.3\% polarization) and a linewidth of $15.6(3)$ mHz.  Such low $^{129}$Xe polarization relative to the Rb polarization at our Rb density ($n_S\approx10^{13}$cm$^{-3}$) could be due to a temperature-dependent wall relaxation mechanism similar to what has been reported in Rb-$^{3}$He cells~\cite{Babcock2006}. We see no signs of quadrupole beating in the $^{131}$Xe signal and believe the absence of first order quadrupole beating is generic to transverse polarization. From free induction decay studies we estimate residual quadrupole effects to be 2 mHz~\cite{Wu1988}. 


In order to construct $\xi$ we need to measure the resonance frequencies $\Omega^a$ and $\Omega^b$. This is most easily accomplished in our system by feedback~\cite{Bechhoefer2005}. We make corrections to each isotope's drive frequency such that the phase as measured by the Rb ($\delta \mp \epsilon_z$) is kept equal to zero. In the high gain limit, the drive frequency for each isotope becomes $\tilde{\omega}_d =\tilde{\Omega}_0\mp i\omega\tilde{\epsilon}_z$. If we evaluate Eq.~\ref{master} with $\Omega$ replaced with $\omega_d$ for each isotope and call it $\xi'$ we find
\be
\tilde{\xi}'={\rho \tilde{\omega}_d^b-\tilde{\omega}_d^a \over 1+\rho}=\tilde{\xi}+i\omega \tilde{\epsilon}_z.
\ee
The computed comagnetometer signal $\xi'$ not only depends on $X_z^a$ and $X_z^b$ but also exhibits a linear dependence on the derivative of $\epsilon_z$. 

Since we would like to measure $\xi$, we need to detect $\epsilon_z$ and subtract its derivative from $\xi'$. A direct in situ measure of $\epsilon_z$ could be achieved by measuring the phase of an ancillary rotating transverse field. However, for experimental convenience, we chose to monitor the sum of the drive frequencies. Although this sum frequency is independent of $\epsilon_z$ it is dominated by drifts in $B_{z0}$. Insofar as $B_{z0}$ is the dominant contribution to $\epsilon_z$, the sum frequency and $\epsilon_z$ should be highly correlated. As such, we constructed $\epsilon_z$ in post processing as follows
\be
\tilde{\epsilon}_z = \tan^{-1}\left({1\over B_w}{\tilde{\omega}_d^a+\tilde{\omega}^b_d \over \gamma^a+\gamma^b}\right).
\ee
We will show that this assumption dramatically improves the comagnetometer's suppression of low frequency magnetic field noise. 

\begin{figure}
\includegraphics[width = 8.6cm]{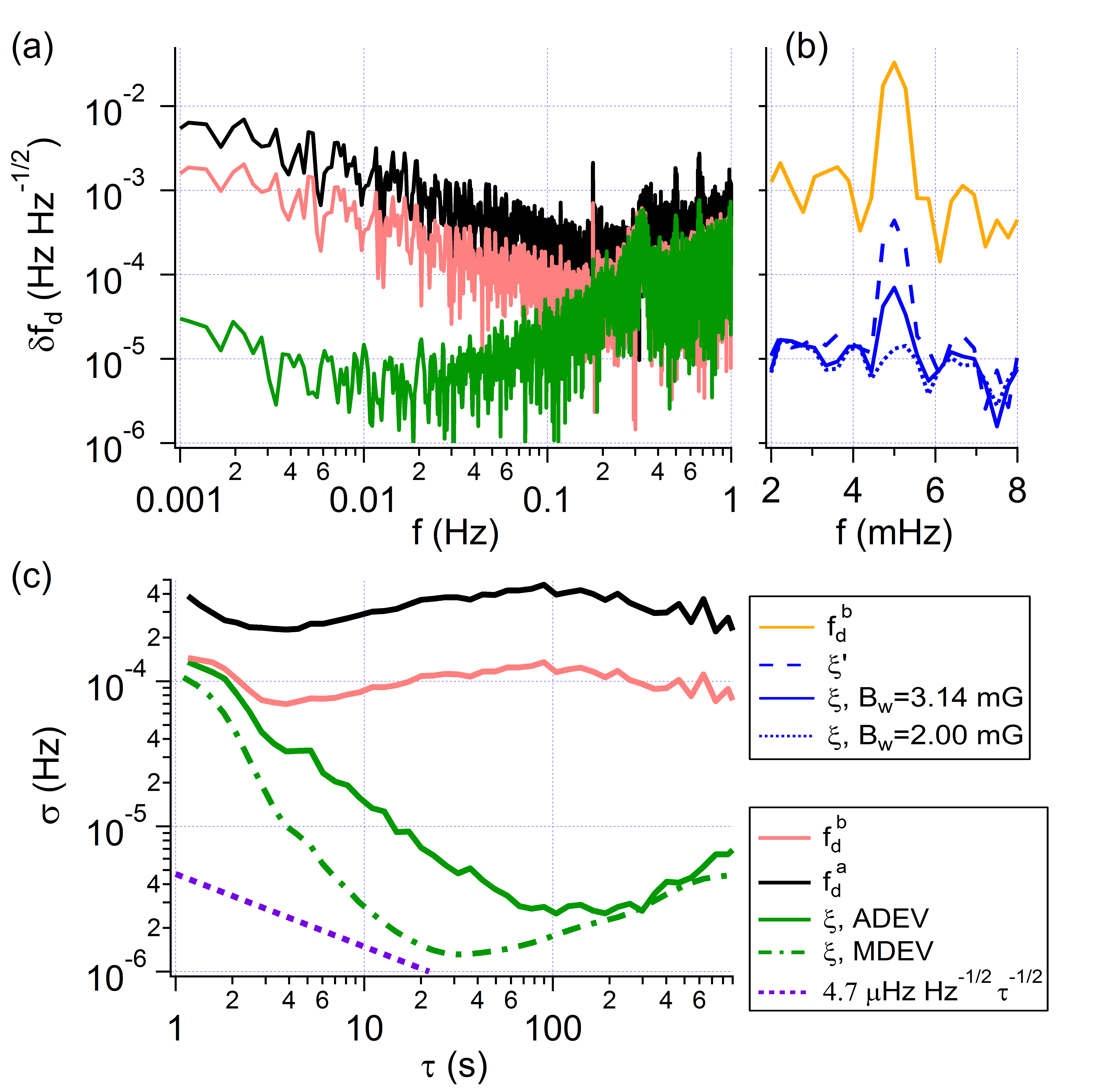}
\caption{ (a) Amplitude spectral densities of $f_d^{b}$, $f_d^{a}$, $\xi$. (b) Spectra with ancillary AC $B_z$ applied showing the field suppression factor for various comagnetometer signal computations. (c) Allan Deviations of $f_d^{b}$, $f_d^{a}$, $\xi$, where $\xi$ was analyzed with both standard and modified Allan deviations. All data shown have been corrected for finite gain. A Hanning window is applied to each spectrum.}\label{adev}
\end{figure}

 We stabilize the two drive frequencies to line center with an accuracy of $\pm0.3$ mHz and record the drive frequencies as a function of time. Once acquired, the individual drive frequencies are corrected for finite gain and used to compute $\epsilon_z$ and $\xi$. Figure~\ref{adev} (a) shows the amplitude spectral densities of $\omega_d^a$, $\omega_d^b$, and $\xi$. The spectrum of $\omega_d^{a}$ is approximately $\rho$ larger than that of $\omega_d^{b}$ from DC to 0.1 Hz, suggesting that $\hat{z}$ magnetic field noise dominates over this frequency band.

In separate experiments, we record the drive frequencies with an ancillary $B_z$ applied. This 5 mHz 4.3 $\mu$G field is used to measure the field suppression factor (FSF), which we define to be $\tilde{\omega}_d^{b}$/$\tilde{\xi}$ at the frequency of the ancillary $B_z$. We find that the FSF depends critically on the assumed value of $B_w$ (see Fig.~\ref{adev} (b)). We measure $B_w$ independently using two methods; from the open loop response of the Xe (see Eq.~\ref{openloop}) we infer $B_w=3.5(3)$ mG, and from the magnetometer's response to $B_y$ we find $B_w = 3.1(1)$ mG. Analyzing the precession data assuming the weighted mean $B_w= 3.14$ mG we find an FSF of 470. However, choosing $B_w = 2.0$ mG produces an FSF of 2300. Alternatively, if we ignore the influence of $\epsilon_z$ completely and calculate $\tilde{\omega}_d^{b}$/$\tilde{\xi}^{\prime}$, we find an FSF of 75. 

In order to suppress uncertainty in the FSF due to uncertainty in $B_w$, we stabilized $B_{z0}$ using the sum $\omega_d^a+\omega_d^b$. By improving the stability of $B_{z0}$ by a factor of 15 at 5 mHz, the FSF increased to 1800. Clearly, $\epsilon_z$ introduces substantial phase shifts in the comagnetometer signal, significantly limiting the FSF if left unaccounted for. 

Figure~\ref{adev}(a) shows $\xi$'s spectrum without an ancillary $B_z$ and without $B_{z0}$ stabilization. $\tilde{\xi}$'s spectrum exhibits white phase noise at high frequency which turns to white frequency noise near $\Gamma_2$. The coefficients of these noise processes are consistent with the finite SNR of our detection. For frequencies lower than 1 mHz $\tilde{\xi}$'s spectrum exhibits random walk of frequency. We find that $\tilde{\xi}$'s spectrum is independent of the $B_w$ used in computation, suggesting that $\xi$'s noise is not limited by the FSF ($B_w=3.14$ mG for the data shown).

Figure~\ref{adev} (c) shows the standard Allan deviation (ADEV) of $\omega_d^{b}$, $\omega_d^{a}$, and $\xi$ as well as the modified Allan deviation (MDEV)~\cite{Allan1981} of $\xi$, each with a 1 Hz step low pass filter applied. The MDEV averages random walk of frequency noise faster than the ADEV and is therefore able to resolve a lower bias instability of roughly 1 $\mu$Hz after 30 seconds of integration. Both the ADEV and MDEV eventually exhibit $\tau^{1/2}$ trends. The source of this frequency drift is uncertain. Possible sources of drift include contributions to $\epsilon_z$ besides $B_z$ (which are not accounted for) and fluctuations in $\Gamma_2$ (which likely stem from Rb density fluctuations and manifest as noise if $\Delta \neq 0$~\cite{Walker2016}).

Fitting the MDEV results~\cite{Vanier1989} in an angle-random walk (ARW) of $\sqrt{2}\times4.7(4)=6.6(6)$ $\mu$Hz/ $\sqrt{\text{Hz}}$. The measured SNRs (5300 and 3200 $\sqrt{\text{Hz}}$ for $^{131}$Xe and $^{129}$Xe, respectively) suggest an ARW limit of 4 $\mu$Hz/$\sqrt{\text{Hz}}$, very similar to that found by fitting the MDEV. The photon shot noise limited ARW is roughly a factor of $10^3$ smaller. It is unclear what is currently limiting the detection noise. Possible noise sources include finite pulse modulation fidelity and imperfect pump-probe laser alignment mapping bias pulse noise onto $S_z$. 

This is the first demonstration of a continuously SE pumped comagnetometer which avoids time-averaged longitudinal alkali-metal and noble gas polarization. The demonstrated field suppression and white frequency noise would be sufficient to realize a bias stability of 200 nHz if stray 1/$f$ magnetic noise were the dominant noise contribution. Despite $30\times$ smaller $b_K^S K_{\perp}$, the ARW is only $6\times$ that of Ref.~\cite{Limes2018} at $230\times$ the bandwidth.

We have shown that the phase of the embedded alkali-metal magnetometer $\epsilon_z$ plays an important role in understanding the comagnetometer. Indeed, the field suppression factor improves dramatically when correcting $B_{z0}$ to keep $\omega_d^a+\omega_d^b=const.$ We note that, in addition to $B_z$, $\epsilon_z$ could also depend on pump pointing, $K_z$, and back polarization ($K_y$ producing $S_y$). By measuring the response of the magnetometer to an ancillary rotating $B_{\perp}$, $\epsilon_z$ could be monitored directly and in real time for greater precision.

The stability demonstrated should be sufficient to make the first measurement of $b_a^b$~\cite{Vaara2019,Limes2019}. Additionally, with a 2 mm cell~\cite{Bulatowicz2013}, we anticipate being able to improve the present upper bound for spin-mass couplings in the sub-mm wavelength range by an order of magnitude after merely 100 seconds of measurement. Future work will include studying the accuracy of the system by measuring Earth's rotation. 

\begin{acknowledgements}
We are grateful to James Pavell for making the vapor cell used in this work. This research was supported by NSF GOALI award numbers PHY-1607439 and PHY-1912543 and by Northrop Grumman Mission Systems' University Research Program.
\end{acknowledgements}

\bibliography{spinexchangecopy}

\end{document}